\documentclass[twocolumn,english,aps,prb,showpacs,superscriptaddress,amssymb,amsfonts]{revtex4}
\usepackage{color}
\usepackage{array}
\usepackage{amsmath}
\usepackage{amssymb}
\usepackage{epsfig}
\usepackage{graphicx}
\usepackage{wasysym}
\usepackage{bm}

\begin{document}

\title{Establishing non-Abelian topological order in Gutzwiller projected Chern insulators via Entanglement Entropy and Modular $\mathcal S$-matrix}

\author{Yi Zhang}
\affiliation{Department of Physics, University of California,
Berkeley, CA 94720, USA}

\author{Ashvin Vishwanath}
\affiliation{Department of Physics, University of California,
Berkeley, CA 94720, USA}

\begin{abstract}
We use entanglement entropy signatures to establish non-Abelian
topological order in projected Chern-insulator wavefunctions. The
simplest instance is obtained by Gutzwiller projecting a filled band
with Chern number $C=2$, whose wavefunction may also be viewed as
the square of the Slater determinant of a band insulator. We
demonstrate that this wavefunction is captured by the $SU(2)_2$
Chern Simons theory coupled to fermions. This is established most
persuasively by calculating the modular $\mathcal S$-matrix from the
candidate ground state wavefunctions, following a recent
entanglement entropy based approach. This directly demonstrates the
peculiar non-Abelian braiding statistics of Majorana fermion
quasiparticles in this state. We also provide microscopic evidence
for the field theoretic generalization, that the N$^{\rm th}$ power
of a Chern number $C$ Slater determinant realizes the topological
order of the $SU(N)_C$ Chern Simons theory coupled to fermions, by
studying the $SU(2)_3$ (Read-Rezayi type state) and the $SU(3)_2$
wavefunctions. An advantage of our projected Chern insulator
wavefunctions is the relative ease with which physical properties,
such as entanglement entropy and modular $\mathcal S$-matrix can be
numerically calculated using Monte Carlo techniques.

\end{abstract}

\maketitle

It is well known that quasiparticles may go beyond the conventional
bosonic and fermionic statistics in two-dimensional many-body
systems. A famous example is the Laughlin quantum Hall state that is
realized by interacting electrons in fractionally filled Landau
levels\cite{FQHEexp}. These are described by the Laughlin
wavefunction\cite{FQHEtheo}, where the quasiparticles carry
fractional statistics. This state realizes an Abelian topological
order \cite{wen1990}, described by a relatively well
understood\cite{Wen1995} low-energy effective theory, the $U(1)$
Chern Simons (CS) theory.

There is increasing interest in generalizations to non-Abelian
statistics, partially brought on by the recent preliminary evidence
for Majorana fermions in superconductor-semiconductor
junctions\cite{majorana2012} and their potential as topological
quantum memories \cite{memory, kitaev2003, nayak}. Other examples of
non-Abelian topological order are the $\nu=5/2$ and $\nu=12/5$
fractional quantum hall effects\cite{tsui1987} and the Moore-Read
states\cite{MooreRead}. Although a general theory and classification
of these states are still absent, there is a wide class of states
effectively characterized by non-Abelian CS theories\cite{wen19913}:
the low-energy effective theory for filling fraction $\nu=k/N$ on
$k$ Landau levels is the $SU(N)_k$ CS theory. In the simplest case
of $SU(2)_2$ CS theory, the elementary excitations are Ising anyons,
whose braiding transforms the ground state, instead of just
incurring phase factors corresponding to anionic statistics as in
the Abelian case\cite{Oshikawa2006}. The entanglement spectrum has
also been established as a powerful tool for identifying such phases
in numerical calculations\cite{HaldaneLi}. For universal quantum
computation, the minimal $SU(2)_2$ non-Abelian statistics is
insufficient. Instead, one needs at least the complexity of
$SU(2)_3$ topological order with Fibonacci anyons \cite{nayak},
which may describe the fractional quantum hall plateau at
$\nu=12/5$. Examples of such phases are given by the Read-Rezayi
states\cite{ReadRezayi} and generalizations to lattice spin liquid
states\cite{Ronny2009, Ronny2011}, however, physical properties of
these constructions are relatively difficult to evaluate.

Fractional quantum Hall liquids are generally associated with
extreme experimental conditions such as clean samples and large
magnetic fields. Yet it is increasingly being appreciated  that
Landau levels are  not the sole route to realizing these states. It
is well known that the integer quantum Hall effect is present in
Chern insulators - lattice band models without an external magnetic
field but with a net Berry curvature in reciprocal
space\cite{Haldane1988}. Analogous interacting lattice models offer
a new route to realizing topological orders, for which there has
been mounting numerical evidence both for Abelian\cite{Bernevig2011,
Chamon2011, donna2011, sheng20112, verderbos} and
non-Abelian\cite{Yao, Bernevig2012a, Bernevig2012, ShuoYang2012}
states, which are collectively referred to as fractional Chern
insulators.

Recall, the Laughlin wavefunctions \cite{FQHEtheo}
$\psi\sim\prod\left(z_i-z_j\right)^m e^{-\left|z_k\right|^2/4}$, can
be considered as the $m^{\rm th}$ power of a lowest-Landau-level
integer Quantum Hall state of anyons with a reduced charge.
Previously, we confirmed that the lattice analog of this statement:
the $m^{\rm th}$ power of a Chern band wavefunction with unit Chern
number $\psi\sim\psi^m_{C=1} $ has the topological order of a
Laughlin state of order $m$\cite{frank2011b, smat}. In this paper,
we focus on the cases when the Chern number $C>1$, which is {\em
unique} to lattice models and has no simple Landau level
counterpart. Consistent with the field theory study in Ref.
\onlinecite{Wen1999} and parton construction scheme proposed in Ref.
\onlinecite{YML2012}, we suggest that the square (power $N=2$) of
$C=2$ Chern band wavefunctions $\psi\sim\psi^2_{C=2} $ are captured
by the $SU(2)_2$ CS theory coupled to fermions and have the same
quasiparticle braiding statistics as the Moore-Read Pfaffian
state\footnote{Strictly speaking this wavefunction is not the same
topological order as the Moore Read Pfaffian state captured by the
pure $SU(2)_2$ CS theory; although it shares the same ground-state
degeneracy and quasiparticle excitation braiding encoded in the
modular $\mathcal S$-matrix with the Pfaffian state, it differs at
the level of the quasiparticle spin encoded in the modular
$T$-matrix due to the fermionic core for certain quasiparticles.
Similar difference exists between the $SU(2)_k$ states we study and
the Read-Rezayi states. See Ref. \onlinecite{Maissam2010} for
detailed discussion. We do not focus on such distinctions because
these states behave identical for the properties we study.}
$\psi\sim
\mbox{Pf}\left(\frac{1}{z_i-z_j}\right)\prod\left(z_i-z_j\right)
e^{-\left|z_k\right|^2/4}$. We verify that there are only three
linearly independent wavefunctions by construction, consistent with
the expected three-fold ground-state degeneracy. Especially, the
wavefunction diagnostic algorithm in Ref. \onlinecite{smat} can be
generalized to non-Abelian cases and particularly useful for
many-body system where entanglement spectrum is not available. With
variational ansatz, physical measurable of these states are much
simpler to calculate, thus we are able to extract the modular
$\mathcal S$-matrix easily and determine the quantum dimension and
quasiparticle statistics through topological entanglement entropy
(TEE)\cite{kitaev2006, levin2006, frank2011b, smat}, and prove the
existence of non-Abelian quasiparticles. To our knowledge, this is
the first direct numerical measurement of the modular $\mathcal
S$-matrix and identification of a non-Abelian topological order
wavefunction. We also generalize our studies of ground-state
degeneracy and entanglement to the $C=2, N=3$
($\psi\sim\psi^3_{C=2}$) and $C=3, N=2$ ($\psi\sim\psi^2_{C=3}$)
cases. These results imply the effective theory for the $N^{\rm th}$
power of a $C=k$ Chern insulator's band $\psi\sim\psi^N_{C=k} $ is
the $SU(N)_k$ CS theory coupled to fermions, allowing non-Abelian
statistics when $N>1$ and $k>1$. Besides, these constructions may
offer access to the entire ground-state manifold with different
choices of boundary conditions of the parent Chern insulator.

{\em \underline{Chern Number $C=2$ Model:}} To construct a two-band
model with Chern number $C=\pm2$, consider a tight-binding model on
a two-dimensional square lattice with two orbitals on each lattice
site labeled by $I=1,2$:
\begin{eqnarray}
H&=&\underset{<ij>,I}{\sum}(-1)^{I+1}c_{jI}^{\dagger}c_{iI}+\underset{<ij>}{\sum}\left(e^{i2\theta_{ij}}c_{j2}^{\dagger}c_{i1}+h.c.\right)
\nonumber \\
&+&\Delta\underset{<<ik>>}{\sum}\left(e^{i2\theta_{ik}}c_{k2}^{\dagger}c_{i1}+h.c.\right)
\label{eqn:ham_k2}
\end{eqnarray}
where $\theta_{ij}$ is the azimuthal angle for the vector connecting
$i$ and $j$. By counting the number of chiral modes on the physical
edges as well as within the entanglement spectrum, we verify that
the model has a finite gap between the two bands with Chern number
$C=\pm2$ respectively. Hereafter, we assume $\Delta=1/\sqrt{2}$ for
a maximum gap to suppress the finite size effect. For a system at
half filling with periodic boundary conditions, the many-body ground
state occupies the valence band below the band gap, and the
corresponding wavefunction $\chi(z_{1},z_{2}\cdots)$ is a Slater
determinant, where $z=\left(\vec r,I\right)$ contains both the
position and orbital indices.

{\em \underline{Gutzwiller projected wavefunctions:}} Our
wavefunctions' construction generalizes previous chiral spin liquid
parton constructions\cite{frank2011b, kalmeyer, kalmeyer1989,
wenzee}, but instead of occupying a band with Chern number $C=1$,
each parton now fills up a band with Chern number $C>1$, e.g. the
Hamiltonian in Eqn. \ref{eqn:ham_k2}. It is then restricted to one
fermion per site Hilbert space by Gutzwiller projection. For the
simplest case with two flavors of partons labeled as spin-up and
spin-down at half-filling, the resulting wavefunction is
$\Phi(z_{1},z_{2}\cdots)=\chi_\uparrow(z_{1},z_{2}\cdots)\chi_\downarrow(\tilde
z_{1},\tilde z_{2}\cdots)$, where $\tilde z_{i}$ are the set of
complementary sites of $z_{i}$. Note that all the charge degrees of
freedom are now projected out, $\Phi(z_{1},z_{2}\cdots)$ is a spin
wavefunction and purely bosonic. In addition, Eqn. \ref{eqn:ham_k2}
has a particle-hole symmetry $c^\dagger_{\vec rI}\leftrightarrow
c_{\vec rI} \cdot (-1)^{r_x + r_y}$, which simplifies
$\Phi(z_{1},z_{2}\cdots)=\chi^2(z_{1},z_{2}\cdots)$ upto an
unimportant sign. The properties of $\chi^2(z_{1},z_{2}\cdots)$ are
the major focus of this paper. Note that it is $\pi/2$ rotational
symmetric even though $\chi(z_{1},z_{2}\cdots)$ is not; although the
$\pi/2$ rotation symmetry is not essential to the topological
properties, it is especially helpful for their
determination\cite{smat}.

To construct $\chi^2(z_{1},z_{2}\cdots)$ with periodic boundary
conditions, there are multiple choices of boundary conditions for
the parent Chern insulator, e.g. either periodic or antiperiodic
boundary condition along the $\hat{x}$ and $\hat{y}$ directions in
Eqn. \ref{eqn:ham_k2}. Let's denote the four corresponding projected
wavefunctions as $|\Phi_{x}\Phi_{y}>$, $\Phi_{x,y}=0,\pi$. Physical
quantities related to the wavefunctions may be calculated with
variational Monte Carlo method\cite{gros1989}. To understand their
relation and linear dependence, we calculate the overlaps between
them with variational Monte Carlo method on a $12\times12$ system:
\begin{eqnarray}
\left\langle 00|\pi\pi\right\rangle &=&\alpha \nonumber \\
\left\langle 0\pi|\pi0\right\rangle &=&\alpha' \nonumber \\
\left\langle 0\pi|00\right\rangle &=&\left\langle
\pi0|00\right\rangle
=\beta \nonumber \\
\left\langle 0\pi|\pi\pi\right\rangle &=&\left\langle
\pi0|\pi\pi\right\rangle =\beta' \label{eqn:overlap}
\end{eqnarray}

Numerically we find to very high accuracy that $\alpha=\alpha'=-0.086$ and $\beta=\beta'=0.457$. We may construct a "generalized" projection operator:
\begin{eqnarray}
P&=&\sum\left|\Phi_{x}\Phi_{y}\right\rangle \left\langle
\Phi_{x}\Phi_{y}\right|\left|\Phi'_{x}\Phi'_{y}\right\rangle
\left\langle \Phi'_{x}\Phi'_{y}\right| \nonumber
\\&=&\left(\begin{array}{c}
\left|\pi0\right\rangle \\
\left|0\pi\right\rangle \\
\left|\pi\pi\right\rangle \\
\left|00\right\rangle
\end{array}\right)^{T}\left(\begin{array}{cccc}
1 & \alpha & \beta & \beta\\
\alpha & 1 & \beta & \beta\\
\beta & \beta & 1 & \alpha\\
\beta & \beta & \alpha & 1
\end{array}\right)\left(\begin{array}{c}
\left\langle \pi0\right|\\
\left\langle 0\pi\right|\\
\left\langle \pi\pi\right|\\
\left\langle 00\right|
\end{array}\right)\nonumber \\
\label{eqn:project}
\end{eqnarray}

Due to the non-orthogonality between the basis states, the
eigenvalues of $P$ actually contains one $0$, so the corresponding
eigenstate is projected out:
\begin{equation}
P\left[|\pi\pi>-|0\pi>-|\pi0>+|00>\right] = 0 \label{eqn:f-1}
\end{equation}
where we have used $2\beta=1+\alpha$ (true to high numerical
accuracy). Thus there are only three linearly independent
wavefunctions by construction:
\begin{eqnarray}
|F_{x}=1,F_{y}=1>&\simeq&\left(|00>+|0\pi>+|\pi0>+|\pi\pi>\right) \nonumber \\
|F_{x}=1,F_{y}=-1>&\simeq&\left(|00>-|0\pi>+|\pi0>-|\pi\pi>\right)\nonumber \\
|F_{x}=-1,F_{y}=1>&\simeq&\left(|00>+|0\pi>-|\pi0>-|\pi\pi>\right) \nonumber \\
\label{eqn:fxfy}
\end{eqnarray}
upto phase and normalization. We have introduced the flux threading
operators $F_{x}$ and $F_{y}$ to label these states. The
wavefunctions' linear dependence is consistent with the ground-state
degeneracy of $SU(2)_2$ CS theory, as shown in the supplementary
material. These provide a complete basis for candidate ground-state
wavefunctions, i.e. other possible constructions with different
boundary conditions for partons are shown to be linearly dependent
on the wavefunctions above\footnote{See supplementary material
online}.

{\em \underline{Topological Entanglement Entropy:}} To obtain
further information on the wavefunctions, we extract their TEE for a
$6\times6$ system following the prescription of Kitaev and
Preskill\cite{kitaev2006}, which effectively cancels out the
boundary and corner contributions and exposes the topological term,
given the size of the regions exceed the correlation length.
Although the smallest length scale is only $2$ lattice spacings for
the system size we study, it is still longer than the correlation
length $\xi\sim0.5$ lattice spacing. In addition, the corresponding
wavefunction overlaps suggest that the residue of Eqn. \ref{eqn:f-1}
is just $\sim 1.3\%$, thus the orthogonal basis in Eqn.
\ref{eqn:fxfy} is still a good approximation. These facts suggest
that the system size is large enough to reflect universal
properties. Throughout we focus on the Renyi entropy $S_2$ due to
its ease of calculation\cite{frank2011, frank2011b, smat}.

We find that the TEE of $|\Phi_{x}\Phi_{y}>$ for a topologically
trivial disc-shape entanglement partition is $\gamma=0.85\pm0.08$,
in reasonable agreement with the ideal theoretical value
$\lambda_{SU(2)_2}=\log2\sim0.693$ for the $SU(2)_{2}$ CS theory.
Note that for an Abelian topological order with $D^{2}=3$ degenerate
ground states on a torus, the expected TEE would be $\gamma=\log
D\sim0.549$, which deviates further from the calculated value and is
unlikely to describe these wavefunctions.

\begin{figure}
\begin{center}
\includegraphics[scale=0.45]{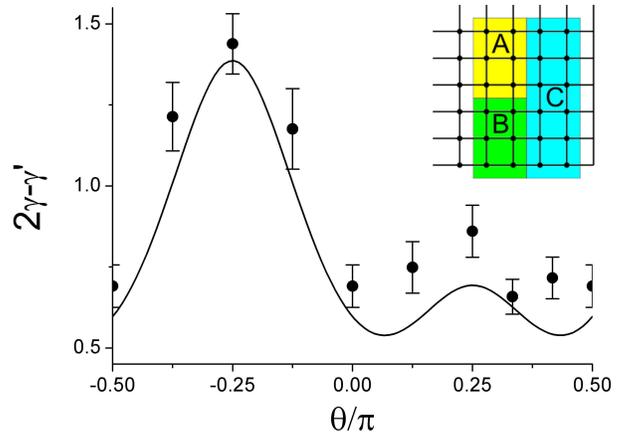}
\caption{(Inset) Kitaev Preskill scheme for extracting TEE by
partitioning the torus into regions A, B and C$\cite{kitaev2006}$.
Regions $C$ and $AB$ encircle the torus, leading to a
state-dependent TEE of $\gamma'$, while the TEE $\gamma$ for region
A and B has a fixed value. The resulting
$2\gamma-\gamma'=-S_{ABC}+S_{AB}+S_{BC}+S_{AC}-S_A-S_B-S_C=-2
S_A+2S_{AB}-S_{ABC}$ is plotted for the linear combinations of
wavefunctions $|\Phi_1>$. The solid curve is the theoretical value
from the $SU(2)_2$ CS theory.} \label{fig1}
\end{center}
\end{figure}

\begin{figure}
\begin{center}
\includegraphics[scale=0.70]{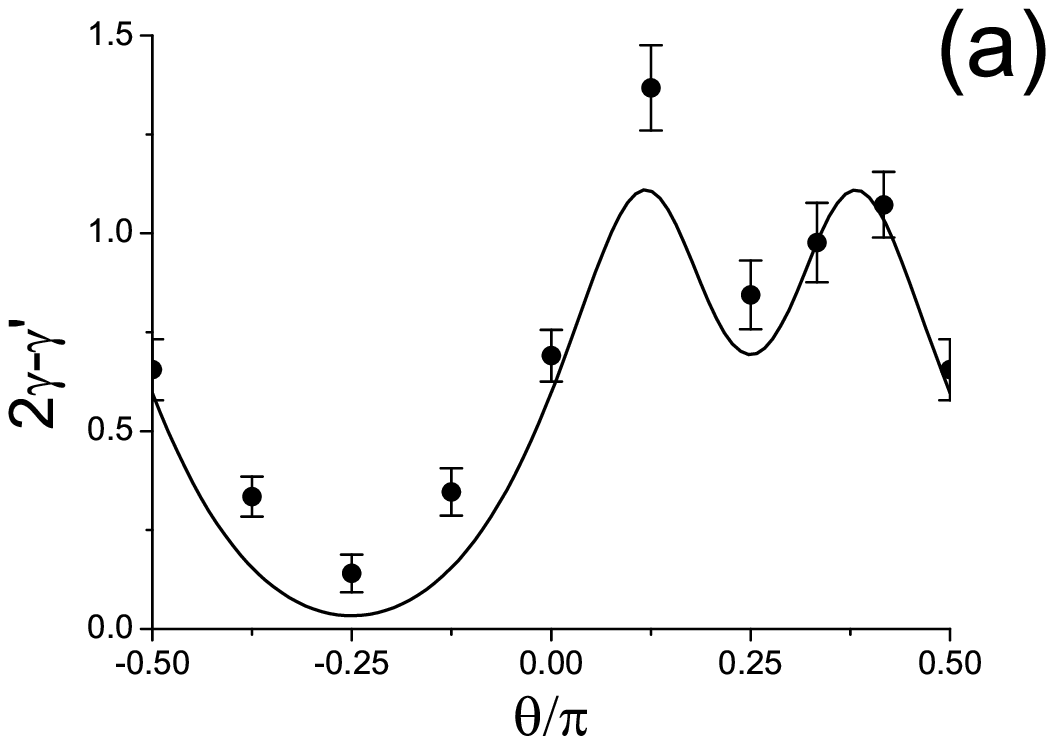}
\includegraphics[scale=0.70]{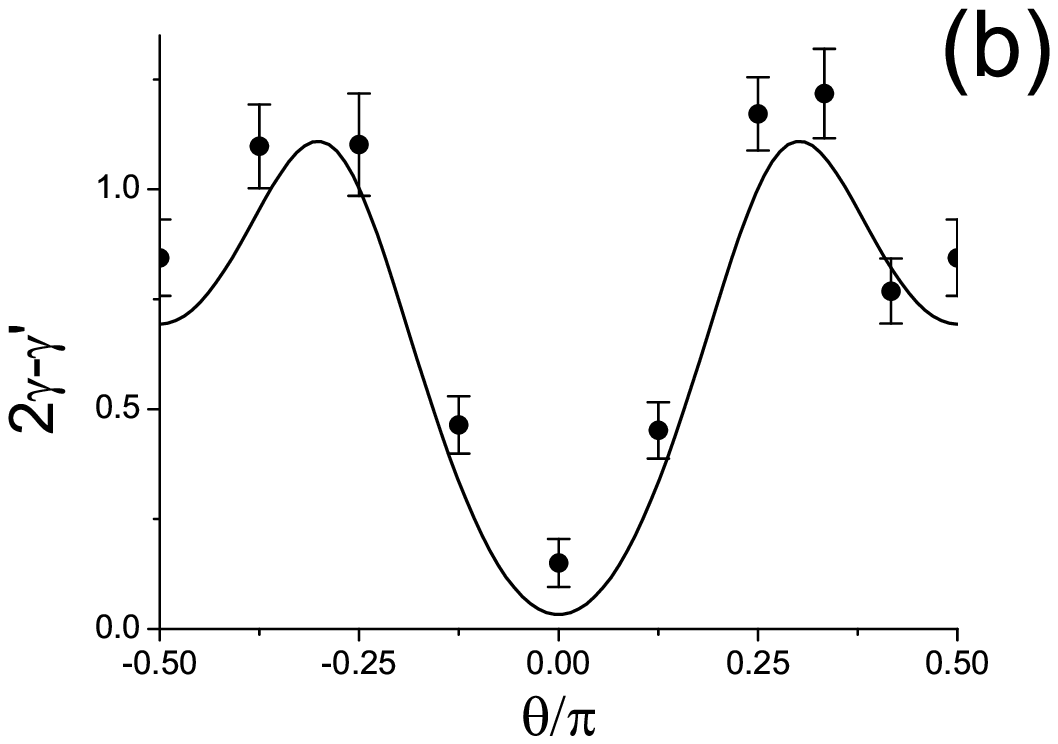}
\caption{TEE $2\gamma-\gamma'$ for linear combinations of
wavefunctions: (a) $|\Phi_2>$, and (b) $|\Phi_3>$. The solid curves
are the theoretical values from the $SU(2)_2$ CS theory. The
presence of only two minima with $2\gamma-\gamma'\simeq 0$ indicates
that two of the three quasiparticles are Abelian, while the third
one must be non-Abelian.} \label{fig2}
\end{center}
\end{figure}

{\em \underline{Modular $\mathcal S$-Matrix from Entanglement
Entropy:}} A decisive identification of the topological order is
provided by extracting the braiding properties of quasiparticles
using entanglement. Following the algorithm for a $\pi/2$ rotation
symmetric system in Ref. \onlinecite{smat}, we (i) calculate the TEE
$\gamma'$ for partitioning the torus into two cylinders along the
$\hat y$ direction for various linear combinations of wavefunctions,
see Fig. \ref{fig1} inset, then (ii) search for the states with
minimum entanglement entropy (maximum TEE $\gamma'$) and identify
them as the $\hat y$ direction Wilson loop states of quasiparticles,
and finally (iii) establish their transformation under $\pi/2$
rotation, which gives the modular $\mathcal S$-matrix. The numerical
results of TEE shown as $2\gamma-\gamma'$ for the following linear
combinations: $|\Phi_1>=\cos\theta|0\pi>+\sin\theta|\pi 0>$,
$|\Phi_2>=\sin\theta|0 0>-\cos\theta|\pi 0>$, and
\begin{eqnarray}
|\Phi_3>&=&(\sin\theta+0.7915\cos\theta)|0 0> \nonumber\\
&-&(\sin\theta+0.4697\cos\theta)|\pi 0> \nonumber\\
&-&1.2623\cos\theta|0 \pi>
\end{eqnarray}
for selected values of $\theta$ are shown in
Fig. \ref{fig1}, Fig. \ref{fig2}a and Fig. \ref{fig2}b,
respectively. By identifying the minima of $2\gamma-\gamma'$, we
obtain the three orthogonal quasiparticle states, given
approximately as\cite{smat}:
\begin{eqnarray}
|\Xi_1> &=& -|\pi 0>-|0 0> \nonumber \\
|\Xi_2> &=& -|\pi 0>+|0 0> \nonumber \\
|\Xi_3> &=& 0.7915|0 0> - 0.4697|\pi 0> -1.2623|0 \pi>
\label{eqn:mes}
\end{eqnarray}

Now consider a rotation of $\pi/2$: the states $|\Phi_{x}\Phi_{y}>$
transforms as $|\pi 0>\leftrightarrow|0 \pi>$ and $|0
0>\leftrightarrow|0 0>$, which determines the transformation of
$|\Xi_j>$. Along with Eqn. \ref{eqn:overlap} to ensure the
orthogonality of wavefunctions, it leads to the following $\mathcal
S$-matrix:
\begin{equation}
\mathcal S=\left(\begin{array}{ccc}
0.627 & 0.610 & 0.484\\
0.610 & 0.000 & -0.792\\
0.484 & -0.792 & 0.372\end{array}\right) \label{eqn:smat}
\end{equation}

As a comparison, the ideal $\mathcal S$-matrix for the $SU(2)_{2}$
CS theory is:
\begin{equation}
\mathcal S=\left(\begin{array}{ccc}
0.5 & 0.707 & 0.5\\
0.707 & 0 & -0.707\\
0.5 & -0.707 & 0.5\end{array}\right)
\end{equation}

While there is a reasonably quantitative agreement, what is more
revealing are the robust qualitative features of quasiparticle
braiding the obtained $\mathcal S$-matrix Eqn. \ref{eqn:smat}
implies. While the quasiparticles corresponding to $|\Xi_1>$ and
$|\Xi_3>$ obey Abelian statistics upon braiding, the zero diagonal
entry in the modular $\mathcal S$-matrix for the quasiparticle of
$|\Xi_2>$ is a signature of its non-Abelian self statistics. Indeed,
for the Majorana fermion in the $SU(2)_{2}$ CS theory, one Dirac
fermion composes a pair of Majorana fermions $c=\gamma_1+i\gamma_2$;
when one of the Majorana fermions braids with another Majorana
fermion and picks an additional $\pi$ phase, it changes the original
annihilation operator to creation operator and vice versa and fails
to return to the excitation-free ground state, thus the
corresponding entry in the modular $\mathcal S$-matrix vanishes.

We make two more detailed comparisons between numerics on these
wavefunctions and the $SU(2)_{2}$ CS theory, which predicts the
following connection between the eigenstates of $F_x$ and $F_y$ in
Eqn.\ref{eqn:fxfy} and the quasiparticles:\cite{Oshikawa2006}:
\begin{eqnarray}
|F_{x}=1,F_{y}=1>&=&(|1_{y}>+|\psi_{y}>)/\sqrt{2} \nonumber \\
|F_{x}=1,F_{y}=-1>&=&(|1_{y}>-|\psi_{y}>)/\sqrt{2} \nonumber \\
|F_{x}=-1,F_{y}=1>&=&|\sigma_{y}> \label{eqn:isinganyon}
\end{eqnarray}
where $|1_{y}>$, $|\psi_{y}>$ and $|\sigma_{y}>$ are the $\hat y$
direction Wilson loop states of the identity, fermionic and
non-Abelian quasiparticles, respectively. Connection between Eqn.
\ref{eqn:fxfy} and Eqn. \ref{eqn:isinganyon} gives the expressions
of the $|\Phi_{x}\Phi_{y}>$ states in the $|1_{y}>$, $|\psi_{y}>$,
$|\sigma_{y}>$ basis:
\begin{eqnarray}
\left|00\right\rangle &=& 0.8466|1_{y}> + 0.1101|\psi_{y}> + 0.5208|\sigma_{y}> \nonumber\\
\left|\pi0\right\rangle &=&  0.8466|1_{y}> + 0.1101|\psi_{y}> - 0.5208|\sigma_{y}> \nonumber\\
\left|0\pi\right\rangle &=&  0.1101|1_{y}> + 0.8466|\psi_{y}> + 0.5208|\sigma_{y}> \nonumber\\
\left|\pi\pi\right\rangle &=&  0.1101|1_{y}> + 0.8466|\psi_{y}> -
0.5208|\sigma_{y}> \label{eqn:wfrelation}
\end{eqnarray}
consistent with Eqn. \ref{eqn:mes}. In addition, for an arbitrary
ground state, the value of TEE is given by\cite{dong2008}:
\begin{equation}
2\gamma-\gamma'=-\log\left(\sum_j p_j^2/d_j^2\right)
\label{eqn:teeformula}
\end{equation}
where $d_j$ and $p_j$ are the individual quantum dimension and the
statistical weight for the $j^{\rm th}$ quasiparticle. We may derive
$d_j$ from the values of the $2\gamma-\gamma'$ minima as
$\log(d^2_j)=2\gamma-\gamma'_j$, which follows straightforwardly
from Eqn. \ref{eqn:teeformula}, and we obtain $d_1=d_\psi=1,
d_\sigma = \sqrt 2$. The values of $2\gamma-\gamma'(\theta)$ with
these individual quantum dimensions and Eqn. \ref{eqn:wfrelation}
are shown in Fig. \ref{fig1} and Fig. \ref{fig2} as the solid
curves, and fit well with the numerical results. In particular,
$d_\sigma=\sqrt 2$ implies that the $\sigma$ quasiparticle must obey
non-Abelian statistics.

{\em \underline{Other non-Abelian States} (i) $SU(3)_2$}: Such
wavefunction constructions may be generalized to even more
complicated non-Abelian cases. As another example, we construct nine
candidate ground-state wavefunctions $|\Phi_{x}\Phi_{y}>$ with
boundary conditions $\Phi_{x,y}=0,\pm2\pi/3$ for
$\chi^3(z_{1},z_{2}\cdots)$, the cube of the Chern insulator
wavefunctions. Repeating the calculations and analysis in previous
sections on a $12\times12$ system, we obtain a "generalized"
projection operator $P'$, see Ref. 41, which has only six non-zero
eigenvalues, therefore there are only six linearly independent
wavefunctions by construction. The $\pi/2$ rotation eigenvalues of
the six corresponding eigenstates are $\pm 1$, $\pm 1$ and $\pm i$.
These results are consistent with the $SU(3)_{2}$ CS theory. In
addition, the TEE for a contractible entanglement partition on a
$6\times 6$ system is $\gamma\simeq1.264\pm0.073$, consistent with
the theoretical value of $D=\sqrt{3(5+\sqrt 5)/2}$ and
$\gamma_{SU(3)_2}=\log D \simeq 1.19$. While we consider these
evidences sufficient, we leave further verifications such as TEE
ground-state dependence and constructions from other boundary
conditions to future works.

{\em (ii) $SU(2)_3$ in close connection to the Read-Rezayi state:}
In analogy to Eqn. \ref{eqn:ham_k2}, we may construct a triangular
lattice tight-binding model with the azimuthal angular dependence
$3\theta$. The model is a two-band Chern insulator with Chern number
$C=\pm 3$ for $\Delta\neq 0$. Similar models may have potential for
the construction of bands with even higher Chern number, and a
systematic scheme to produce arbitrary Chern number bands has been
studied in Ref. \onlinecite{ShuoYang2012}. We construct
wavefunctions with nine different boundary conditions on a
$12\times12$ system. Our results show that only four of the nine
eigenvalues of the corresponding "generalized" projection operator
are unambiguously finite, consistent with the four-fold
ground-states degeneracy of the $SU(2)_{3}$ CS theory.

In conclusion we have introduced lattice wavefunction constructions
for a class of non-Abelian topological phases that (i) are readily
generalized to capture $SU(N)_k$ topological order and (ii) easily
generate the set of candidate ground-state wavefunctions
corresponding to the topological degeneracy and (iii) can compute
physical properties with Monte Carlo techniques, and the usefulness
of entanglement in the diagnosis and study of these wavefunctions.
The presence of such natural lattice wavefunctions holds promise
that such states may be realized in the context of fractional Chern
insulators.

{\bf Acknowledgements:}

We thank Yuan-ming Lu, Tarun Grover, Ying Ran and Maissam Barkeshli
for helpful discussions. This work is supported by NSF DMR-1206728.

\def\urlprefix{}
\def\url#1{}
\bibliographystyle{apsrev}
\bibliography{bibliography}

\end{document}